\documentclass[letterpaper]{article} % DO NOT CHANGE THIS
\usepackage{aaai24}  % DO NOT CHANGE THIS
\usepackage{times}  % DO NOT CHANGE THIS
\usepackage{helvet}  % DO NOT CHANGE THIS
\usepackage{courier}  % DO NOT CHANGE THIS
\usepackage[hyphens]{url}  % DO NOT CHANGE THIS
\usepackage{graphicx} % DO NOT CHANGE THIS
\usepackage{natbib}  % DO NOT CHANGE THIS AND DO NOT ADD ANY OPTIONS TO IT
\usepackage{caption} % DO NOT CHANGE THIS AND DO NOT ADD ANY OPTIONS TO IT
\usepackage{algorithm}
\usepackage{algorithmic}
\usepackage{multirow}
\usepackage{tabularx}
\usepackage{subcaption}
\usepackage{amsthm}
\usepackage{amssymb}
\usepackage{amsmath}
\theoremstyle{definition}
\newtheorem{definition}{Definition}
\usepackage{booktabs}

\usepackage{newfloat}
\usepackage{listings}
\DeclareCaptionStyle{ruled}{labelfont=normalfont,labelsep=colon,strut=off} % DO NOT CHANGE THIS
\floatstyle{ruled}
\newfloat{listing}{tb}{lst}{}
\floatname{listing}{Listing}

\title{
\centering
MDGNN: Multi-Relational Dynamic Graph Neural Network for Comprehensive and Dynamic Stock Investment Prediction
}
\author{Hao Qian\textsuperscript{\rm 1}, 
    Hongting Zhou\textsuperscript{\rm 1}, 
    Qian Zhao\textsuperscript{\rm 1}, 
    Hao Chen\textsuperscript{\rm 1}, 
    Hongxiang Yao\textsuperscript{\rm 2},\\ 
    Jingwei Wang\textsuperscript{\rm 1}, 
    Ziqi Liu\textsuperscript{\rm 1}, 
    Fei Yu\textsuperscript{\rm 1}, 
    Zhiqiang Zhang\textsuperscript{\rm 1}, 
    Jun Zhou\textsuperscript{\rm 1}\footnote{Corresponding author}}
\affiliations{
    \textsuperscript{\rm 1}Ant Group,Hangzhou,China\\
    \textsuperscript{\rm 2}Alibaba Group,Hangzhou,China\\
\{qianhao.qh,zhouhongting.zht,zq317110,chuhu.ch,wangjingwei.wjw,ziqiliu,lingyao.zzq,jun.zhoujun\}@antgroup.com \\
    henry.yhx@alibaba-inc.com
}
\begin{document}

\maketitle

\begin{abstract}
The stock market is a crucial component of the financial system, but predicting the movement of stock prices is challenging due to the dynamic and intricate relations arising from various aspects such as economic indicators, financial reports, global news, and investor sentiment. Traditional sequential methods and graph-based models have been applied in stock movement prediction, but they have limitations in capturing the multifaceted and temporal influences in stock price movements. To address these challenges, the Multi-relational Dynamic Graph Neural Network (MDGNN) framework is proposed, which utilizes a discrete dynamic graph to comprehensively capture multifaceted relations among stocks and their evolution over time. The representation generated from the graph offers a complete perspective on the interrelationships among stocks and associated entities. Additionally, the power of the Transformer structure is leveraged to encode the temporal evolution of multiplex relations, providing a dynamic and effective approach to predicting stock investment. 
Further, our proposed MDGNN framework achieves the best performance in public datasets compared with state-of-the-art (SOTA) stock investment methods.
\end{abstract}

%%%%%%%%%%%%%%%%%%%%%%
\section{Introduction}
The stock market is a crucial component of the financial system, offering investors a marketplace to trade shares of a wide range of assets. Nevertheless, predicting the movement of stock prices is challenging due to the dynamic and intricate relations arising from various aspects. The active trading behaviors of investors, such as buying and selling, drive the fluctuations in stock prices. Additionally, the stock market is influenced by several factors, including economic indicators, financial reports, global news, political events, investor sentiments, and many others. 
Hence, it's indispensable to integrate comprehensive and multifaceted relations to capture the dynamics of the stock markets accurately.

Two lines of research have been applied to stock movement prediction. Traditional sequential methods~\cite{lstm,gru,alstm,tra,sfm} propose to capture the temporal patterns of stock movement by optimizing the temporal dependency encoder, which employs sequential extraction techniques~\cite{rnn, devlin2018bert}. 
Nevertheless, the majority of these methods still assume that stocks are independent of each other and overlook the influence of complex relations. 
In addition, graph-based models~\cite{hist,mrgn,HATR,STHAN-SR,ALSP-TF} incorporate heterogeneous information explicitly from data or implicitly mine it from textual data to capture the interdependence of stocks by designing various graph representation methods. 
However, these approaches could still be dissatisfactory due to the following two issues.

(1) \textbf{Multifacetedness}. 
The stock price movement is influenced not only by a single factor but also by multiple relations among stocks, industries, investment banks, etc. 
For example, changes in the stock prices in a particular industry can be caused by a variety of factors, such as high product demand, new government policies, the rise of raw material costs, and negative earnings reports from large companies. 
Similarly, investment banks can influence stock prices in numerous ways, including conducting research on companies, releasing positive or negative reports, and trading shares. Therefore, accurately predicting the movement of stock prices requires consideration of the multifaceted relations among stocks. Previous graph-based methods for stock investment prediction have only utilized single relations between stocks, ignoring the potential of incorporating other complex relations as auxiliary information.

(2) \textbf{Temporal}.  
The movement of stock prices and the multifaceted relations among stocks are not static but exhibit temporal evolution. Stock prices can change rapidly due to external factors such as economic conditions, political events, and regulatory changes, while internal factors such as company earnings and industry performance can also influence the movement of stock prices over time. 
Relationships among stocks change over time due to factors such as investment banks trading stocks, common shareholders co-holding stocks, and companies releasing products into new industries.
Hence, accurately predicting the movement of stock prices and anticipating the impacts of these changes requires a dynamic approach that considers the historical trends and evolving relationships among stocks. 

To address the aforementioned issues, we introduce a novel framework to underline the multifacetedness and temporal influences in stock investment prediction and propose a \underline{\textbf{M}}ulti-relational \underline{\textbf{D}}ynamic \underline{\textbf{G}}raph \underline{\textbf{N}}eural \underline{\textbf{N}}etwork (\textbf{MDGNN}).
Overall, we utilize the discrete dynamic graph framework to tackle the stock investment prediction. Specifically, to comprehensively capture the multifacetedness nature of stocks, we construct each graph snapshot with daily stock information and relationship data, which is then analyzed with a multi-relational graph embedding layer. 
The generated representation from the multi-relational graph offers a thorough and complete perspective on the interrelationships among stocks and associated entities.
Additionally, we leverage the power of the Transformer structure to encode the temporal evolution of multiplex relations, providing a dynamic and effective approach to predicting stock investment.
Our contributions are summarized as follows:
% \vspace{-0.2em}
\begin{itemize}
\item We discuss the multifacetedness and temporal in the context of stock investment prediction tasks.
We also provide insights on modeling complex stock relations based on empirical evidence.
\item We propose to capture the multifaceted and temporal evolution nature of stocks with a multi-relational dynamic graph and generate a comprehensive representation of the stock market. 
\item We perform extensive experiments on public datasets to verify the superiority of our proposed framework. With detailed analysis, we demonstrate the effectiveness of the multi-relational dynamic graph in tackling the stock investment prediction task.
\end{itemize}

%%%%%%%%%%%%%%%%%%%%%%
\section{Related Work}
\noindent\textbf{Stock Trend Prediction.}
In quantitative trading, the ability to anticipate stock trends is crucial. To accomplish this task, a multi-factor model is commonly employed, as detailed in Nagel's recent work~\cite{nagel2021machine}. This model considers several influential factors from an econometrics standpoint, including trading volumes and prices, as well as company-specific fundamental data like earnings and debt ratio. 
%The obtained prediction results can then be utilized for long/short trading in the real market through either rule- or learning-based methods.

When utilizing learning-based methods, it's a common practice to start with linear regression~\cite{gu2020empirical}. Moreover, ~\cite{roy2015stock} utilized ordinary least squares equipped with regularization, such as ridges and lasso, to overcome the over-fitting issues. However, linear models have limitations in capturing complicated patterns in stock price trends. To overcome this limitation, attempts have been made to incorporate more complex learning techniques. XGBoost~\cite{han2023machine} based method is developed and evaluated through an empirical analysis of companies listed on the NASDAQ. Neural-network-based LSTM~\cite{nelson2017stock} is utilized in predicting future trends of stock prices and shows potential to tackle the challenge of an immensely complex, chaotic, and dynamic environment for the stock market.

%\noindent\textbf{Multi-relational Graph Neural Networks.}
\noindent\textbf{Dynamic Graph Neural Networks.}
The above-mentioned methods for predicting stock trends primarily concentrate on individual stocks and disregard the interdependence and resulting interactions between various stocks. For instance, stocks that belong to the same supply chain are interrelated due to profit transmission. 
%Other types of relations comprise being held by identical institutions, mentioned in the same news article, and so forth.

In recent years, GNN has gained great success owing to its powerful capability of representing complex relations. Traditional GNN methods (e.g., GCN~\cite{gcn}, GAT~\cite{gat}) are mostly based on static graphs where nodes and edges don't change over time. However, many real-world relations (e.g., financial transactions, social relations) are continuously evolving, in which dynamic graphs are indispensable to capture the advancing relations. 
Several approaches have been proposed to represent dynamic graphs. 
%Recently, some studies have started utilizing relationship information and graph learning to predict stock trends. 
One method is RSR~\cite{feng2019temporal}, which incorporates sector and supply chain relation information into its temporal graph convolution. 
%This allows RSR to handle the impact between different stocks by encoding stock relations in a time-sensitive way. 
Another approach is MGRN~\cite{chen2021graph}, which utilizes more relationships such as historical price. This is calculated by the correlation coefficient of two stocks' daily return time series. The multi-graph embedding combined with text embedding extracted from the news is then fed into an LSTM network to predict the stock trend. HATR~\cite{wang2021hierarchical} takes this further by introducing topicality associations in graph modeling. Additionally, Concept-oriented shared information for stock trend forecasting~\cite{xu2021hist} proposed to mine hidden relations by designing a hidden concept module. This approach successfully mined information beyond that carried by predefined concepts. Evolvegcn~\cite{evolvegcn} exploits the combination of graph convolution and RNN~\cite{rnn} to capture both the topological structures and temporal relations. 

\begin{figure*}[ht]
\centering
\includegraphics[width=1.9\columnwidth]{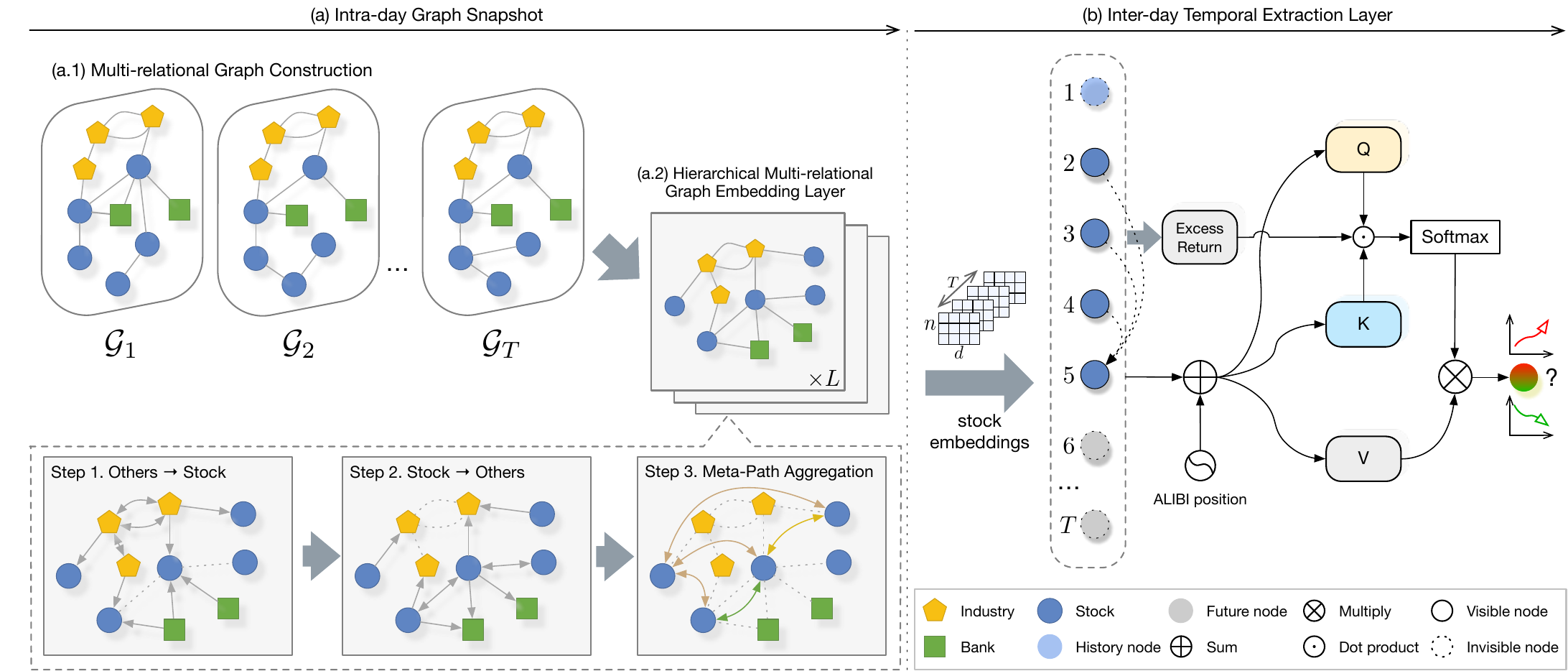}
% \vspace{-0.2cm}
\caption{The overview architecture of the MDGNN Model.}
\label{fig:eldm_model}
 % \vspace{-0.5cm}
\end{figure*}

%%%%%%%%%%%%%%%%%%%%%%
\section{Preliminary}
% \begin{definition}
% {\rm 
% \textbf{Multi-relational Graph.}
% A multi-relational graph is defined as a tuple $\mathcal{G} = (\mathcal{V}, \mathcal{E}, \mathcal{R})$, where $\mathcal{V}$ is the set of nodes, $\mathcal{R}$ is a set of relations, and $\mathcal{E} \subseteq \mathcal{V} \times \mathcal{V} \times \mathcal{R}$ is a set of directed edges. 
% %Each edge is a triplet denoted as $(u,v,r)$ representing that node $u$ and $v$ in relation $r \in \mathcal{R}$.
% Each edge is denoted as a triplet $(u,v,r)$, indicating a relation $r \in \mathcal{R}$ between nodes $u$ and $v$.
% }
% \end{definition}

% \subsection{Problem Formulation}
\begin{definition}
{\rm 
\textbf{Problem Formulation.}
Given that the relationships between stocks are multifaceted and changing on a daily basis, we propose a Dynamic Graph Neural Network (DGNN) to capture and represent them. Let $\mathcal{G}=\{\mathcal{G}_1, \mathcal{G}_2,...,\mathcal{G}_T\}$ represent DGNN, where $\mathcal{G}_t = (\mathcal{V}_t, \mathcal{E}_t, \mathcal{R}_t)$ is a multi-relational graph snapshot at trading day t and $T$ is the total number of snapshots. 
%Concretely, the node set $\mathcal{V}^t$ is composed of the set of stocks, $\mathcal{E}^t$ is the edge set that indicates the correlation intensity between $v_i$ and $v_j$ under a given relation.
For a stock node $v_{it} \in \mathcal{V}_t$, the closing price at a trading day t is denoted as $p_{it}$. 
The ground-truth label of stock $v_{it}$ on trading day t based on the return between two consecutive trading days is denoted as $y_{it} = \frac{p_{i,t+1} - p_{it}}{p_{it}} - {\rm benchmark}_t$, in which ${\rm benchmark}_t$ is the return of the benchmark index on trading day $t$.
%Incorporating information from news and articles related to stocks, we supplement our analysis of the changing relationships between stocks with an event-triggered knowledge graph ($\mathcal{KG}$). 

We formulate the stock prediction as a node regression task that utilizes DGNN to learn a scoring function, $\mathbf{f}(\mathcal{G};\mathbf{\Theta})$, parameterized by $\mathbf{\Theta}$. 
The scoring function is usually optimized by minimizing the loss function as:
\begin{equation}
    \centering
  \mathcal{L} = \sum_{\mathcal{N}} \ell \{\mathbf{Y}, \mathbf{f}(\mathcal{G};\mathbf{\Theta)} \},
   \label{eq:re_rank_define}
\end{equation}
where $\mathcal{N}$ is the set of training samples, and $\ell$ is the loss computed from each sample.
}
\end{definition}

%%%%%%%%%%%%%%%%%%%%%%
\section{Algorithm Design}
%\subsection{Framework Overview}
In the following sections, we will describe the architecture of the MDGNN model as depicted in Figure~\ref{fig:eldm_model}, which includes the Intra-day layer, the Inter-day Temporal Extraction layer, and the prediction layer. 

\subsection{Intra-day Graph Snapshot}
In this section, we outline the framework for capturing node representations from graph snapshots generated on each trading day. The framework comprises two key components: the construction of the multi-relation graph and the graph embedding layer.

\noindent \textbf{Multi-relational Graph Construction}
The performance of a single stock is influenced by a wide range of factors beyond its individual characteristics.
As the stock markets are complex and multifaceted, a singlet relation is not sufficient to depict the intricate relations. 
As such, it is important to consider the correlations between stocks from comprehensive relations so that a more accurate picture of the overall performance of stock markets can be depicted.

To tackle the intricacy of stock markets, we integrate relations from industry, investment banks, and stock pairs to establish a multi-relational graph. This approach enables us to reveal these complex connections and offer a deeper understanding of the underlying dynamics of the financial system.

(1) \emph{Industry Graph}: 
%Industry is important to stocks as it affects the revenue and expenses, etc. 
The performance of a company and its corresponding stock is closely tied to the industry in which it operates. 
For instance, as an industry is growing rapidly, companies in that industry are likely to experience increased demand for their products or services, which will lead to higher revenue and profits. As a result, the stock prices of these companies are likely to increase. 
Besides, the products manufactured by a company can either serve as raw materials for another industry or depend on the raw materials produced by another industry. 
Therefore, any increase in the cost of raw materials can result in an increase in the expenses of companies.
Additionally, government regulations and policies toward industries can significantly impact the associated companies. 
To be specific, we represent the stock as $\mathcal{S}$ and the industry as $\mathcal{I}$, while the connection between them is denoted as $\mathcal{E}_{\mathcal{S}\mathcal{I}}$. This connection contains features that encode the aforementioned supply, demand, competition, and regulatory connections to account for the impact transmission from the industry.

% \vspace{-0.2em}
(2) \emph{Investment Bank Graph}: 
Investment banks greatly impact stock because they provide a wide range of services related to stock markets.
Investment banks often act as market makers for stocks, meaning they provide liquidity to the market by buying and selling stocks on a regular basis. In addition, investment banks provide research reports on stocks  regarding the company's financial performance, industry trends, and other factors. These relationships allow investment banks to have an impact on the price of the stock.
We extract the buy, sell, research, and advisory relations from investment banks to capture the wield significant influence over stock prices.
We denote the investment bank as $\mathcal{B}$ and the connection between that and stock as $\mathcal{E}_{\mathcal{S}\mathcal{B}}$, which incorporates the intricate aforementioned relations.

(3) \emph{Stock Graph}:
Stocks have a great impact on other stocks because of the interconnectedness of the stock market and the various factors that can affect stock prices.
A company's earnings can lead to increased or reduced demand for its stock, as well as other companies in the same industry or sector.
In some cases, companies can be held by the same owners, in which the co-holding relations offer a means of gauging the correlation among stocks.
Additionally, common shareholders may engage in simultaneous buying or selling of a company's stock during a specific period of time.
Therefore, the performance of stocks can be positively or negatively correlated with one another.
To capture the interconnected nature of the stock market, we identify relationships between stocks based on factors such as sector, ownership, and co-holding relations. These relationships are denoted as $\mathcal{E}_{\mathcal{S}\mathcal{S}}$.

To create the multiplex relations mentioned earlier, we start by gathering daily trading data and textual data such as macroeconomic reports, financial news, financial statements, and research reports from TuShare~\footnote{https://tushare.pro/}. 
We then employ financial lexicons and syntactic methods, as suggested by~\cite{Wang_Wang_Li_2020}, to build the edges between pairs of entities.  
Through intricate data preparation and extraction processes, we construct a multi-relational graph that integrates stocks, industries, and investment banks as nodes and multiplex relations as edges.

\noindent\textbf{Hierarchical Multi-relational Graph Embedding Layer}
As stated above, we construct a multi-relational graph from different relationships associated with stocks. This enables us to capture complex representations of the relationships between stocks, leading to a more comprehensive understanding of stock investment modeling. 
Concretely, we define a few meta-paths starting from stock nodes, such as ``Stock-Stock ($\mathcal{S}\mathcal{S}$)", ``Stock-Bank-Stock ($\mathcal{S}\mathcal{B}\mathcal{S}$)", and ``Stock-Industry-Industry-Stock ($\mathcal{S}\mathcal{I}\mathcal{I}\mathcal{S}$)".
As both nodes and edges have a distinct impact on stock nodes, we propose a hierarchical graph embedding layer that can aggregate and propagate information. Besides, edge features are crucial in graph-based models as they encode essential information about the relationships between nodes. For instance, in the context of stock market prediction, the features that encode supply, demand, competition, and regulatory connections between an industry and its corresponding stocks can provide valuable insights into the future trends of stocks. 

Concretely, we utilize an attention mechanism when aggregating information from neighborhood nodes, allowing the model to attend to distinctive attributes of the edges and the nodes they connect. As demonstrated in ~\cite{gat}, using multi-head attention in the graph attention mechanism is advantageous. Hereby, the attention weight of the $k$-th head when aggregating the neighbors of target node $v_i$ is conducted as follows:
\begin{equation}
\centering
\begin{split}
\beta_{ij} &= a^T [\mathbf{W} \mathbf{h}_i||\mathbf{W}\mathbf{h}_j||\mathbf{W}\mathbf{e}_{ij}] , \\
\alpha_{ij}^k &= \frac{{\rm exp} ( {\rm LeakyReLU} (\beta_{ij}))}{\sum_{j^{\prime}\in \mathcal{N}_i} {\rm exp} ({\rm LeakyReLU} (\beta_{ij^{\prime}}))} ,
\end{split}
% \alpha_{ij}^k = \frac{{\rm exp} ( {\rm LeakyReLU} (a^T [\mathbf{W} \mathbf{h}_i||\mathbf{W}\mathbf{h}_j||\mathbf{W}\mathbf{e}_{ij}]))}{\sum_{j^{\prime}\in \mathcal{N}_i} {\rm exp} ({\rm LeakyReLU} (a^T [\mathbf{W}\mathbf{h}_i||\mathbf{W}\mathbf{h}_{j^{\prime}} ||\mathbf{W}\mathbf{e}_{ij^{\prime}} ]))} ,
\label{eq:gat}
\end{equation}
where $\mathcal{N}_i$ denotes the neighborhood nodes of the target node $i$, $||$ represents the concatenation operation, and $\mathbf{W}$ is the shared projection matrix. Moreover, we aggregate representations from multiple heads and use average pooling to update the target node's representation as follows:
\begin{equation}
\centering
\mathbf{h}_i = \sigma \left( \frac{1}{K} \sum_{k=1}^{K}\sum_{j \in \mathcal{N}_i} \alpha_{ij}^k \mathbf{W}^k \mathbf{h}_j \right), 
\label{eq:gat1}
\end{equation}
where $\mathbf{W}^k$ is the shared projection matrix and $K$ is the total number of heads.
Consequently, we obtain the stock representations from each meta-path. Specifically, we define the stock representations from the meta-paths ``$\mathcal{S}\mathcal{S}$", ``$\mathcal{S}\mathcal{B}\mathcal{S}$", and ``$\mathcal{S}\mathcal{I}\mathcal{I}\mathcal{S}$" as $\mathbf{h}_{i1}$, $\mathbf{h}_{i2}$, and $\mathbf{h}_{i3}$. 

However, combining these representations effectively can be a challenging task. 
Attention mechanisms provide a solution by allowing the model to selectively focus on the most relevant representations for the target node. By assigning different attention weights to each representation, the model can effectively combine and aggregate the information from multiple meta-paths. This process not only captures the most critical information but also helps reduce the noise and redundancy in the representations.

Specifically, we design a relation-aware graph module that aggregates node and relation features from a multi-relational graph in an adaptive manner as follows:
% \vspace{-0.4em}
\begin{equation}
\centering
 \mathbf{h}_{v_i} = \sigma  ( \sum_{j=1}^{3} {\rm Softmax}(\mathbf{W} \mathbf{h}_{ij}) \mathbf{h}_{ij} ), 
\label{eq:relation_attn}
\end{equation}
% \vspace{-0.2em}
where $\mathbf{W}$ is a learnable matrix and $\mathbf{h}_{v_i}$ is the representation of node $v_i$ after incorporating multiple relations among nodes. 
By assigning different attention weights to each representation, the model can prioritize the most informative representations, enhancing the model's ability to capture important patterns and relationships. Moreover, analyzing these weights makes it possible to understand the importance of each relation or meta-path in the final prediction. This interpretability is crucial for understanding the reasoning behind the model's decision-making process.

To enhance the representations of stock nodes, we stack multiple hierarchical multi-relational graph embedding layers. The first layer captures local information, while subsequent layers capture increasingly global information. Hence, as we stack multiple graph layers, nodes that are distant from the originating node will be impacted. This facilitates the modeling of the intricate relationships involved in the transmission of stock information. This also enables the model to learn complex patterns and relationships in the graph, leading to improved performance. 
Specifically, we stack L GNN layers to obtain the representation of each stock node $v$ on a trading day $t$, denoted as $\mathbf{h}_{vt}$, using the final GNN layer.

% \vspace{-0.2em}
\subsection{Inter-day Temporal Extraction Layer}
Although the node representation is obtained through the graph embedding layer from each graph snapshot, the semantics of stocks and the relationships between them are constantly evolving. 
For instance, every day, securities companies adjust their positions by selling the stocks of a company purchased the previous day while buying stocks of companies that have not been purchased. 
Furthermore, the features of stocks (e.g., the momentum, volatility, and yield factors) are also changing due to the impact of market changes every day.
Therefore, it is essential to capture the dynamic nature of nodes and the evolving relations among graph snapshots in the temporal order. 

To tackle the aforementioned challenge, we have developed a temporal extraction module that employs the transformer structure~\cite{NIPS2017_7181}.
This module enables the extraction of the temporal evolution of graph snapshots' propagation by taking in the representation of the target nodes within a time window. 

Concretely, let $\mathbf{H}_{v, t-\delta_t:t} = \{ \mathbf{h}_{vt^{\prime}} | t-\delta_t \le t^{\prime} \le t \}$ represent the node's representation, from trading day t and looking back to the preceding $\delta_t$ trading days. Hereby, the window size $\delta_t$ is a hyperparameter. To facilitate with the structure, we transform $h_{v, t-\delta_t:t}$ into query $\mathbf{Q}$ and key $\mathbf{K}$, and value $\mathbf{V}$ as follows: 
\begin{equation}
    \centering
\begin{split}
    \mathbf{Q = W_Q \mathbf{H}_{v, t-\delta_t:t} }, \\
    \mathbf{K = W_K \mathbf{H}_{v, t-\delta_t:t} }, \\
    \mathbf{V = W_V \mathbf{H}_{v, t-\delta_t:t} }, 
   % \mathbf{Q, K, V = W_Q \mathbf{H}_{v, t-\delta_t:t}, W_K \mathbf{H}_{v, t-\delta_t:t}, W_V \mathbf{H}_{v, t-\delta_t:t} },
 \end{split}
 \label{eq:lln}
\end{equation}
where $\mathbf{W_Q}$, $\mathbf{W_K}$, and $\mathbf{W_V}$ are trainable weight matrices of query, key, and value, respectively. 

In addition, the temporal dependency of graph snapshots is crucial for modeling the dependency in the node's representation. 
Over time, the historical relationship between stocks and current price changes will diminish, making stock prices more susceptible to the influence of recent events. Hereby, we leverage the relative position method proposed in ALIBI~\cite{alibi} that adds a static, non-learnable bias to the query-key dot product. It introduces an inductive bias in favor of recent events, as it imposes a penalty on attention scores between distant query-key pairs. Moreover, the penalty increases in proportion to the distance between a key and a query.

Moreover, we also employ the forward mask to prevent positions in the input sequence from attending to subsequent positions during the self-attention mechanism. It's applied to the attention mechanism's softmax operation to mask out the future positions, ensuring that each position can only attend to the previous positions. 

We calculate the dot product of query and key vectors to capture information between any node pair via a self-attention network, in which the multiplicative operation efficiently captures complex feature interactions.
Then we apply the softmax function to scale the attention weight before multiplying it with the corresponding value vector. 
\begin{equation}
\centering
\mathbf{Z} = {\rm softmax}(\frac{\mathbf{Q}\mathbf{K}^T}{\sqrt{d}}  + m \cdot \mathbf{P} + \mathbf{M}) \mathbf{V}, 
\label{eq:qk}
\end{equation}
where $m$ is a slope parameter, $\mathbf{P}$ is the position bias introduced from ALIBI, and $\mathbf{M}$ is the forward mask matrix. 
$\mathbf{Z}$ combines the significant evolving patterns extracted from the historical data between time period $t-\delta_t$ and $t$. $\mathbf{z}_{vt}$ denotes the representation of stock node $v$ on trading day $t$.

\subsection{Prediction Layer}
In stock investment prediction, we aim to estimate the probability $\hat{y}_{vt}$ that a given stock will yield a positive return on trading day $t$ based on the stock's representation $\mathbf{z}_{vt}$ as:
\begin{equation}
\centering
\hat{y}_{vt} = \sigma (\mathbf{W}_1 \mathbf{z}_{vt}  + \mathbf{b}_1).
\label{eq:pred}
\end{equation}
where $\mathbf{W}_1$ and $\mathbf{b}_1$ are trainable matrices and bias; $\sigma$ is the sigmoid activation function.

\section{Experiments}
To validate the efficacy of our method, we conducted extensive experiments using Chinese stock market data. 
%This section presents a comprehensive overview of our experimental setup, including the comparison baseline, dataset, features, graph relationships, and evaluation metrics. We then showcase the primary experimental results, followed by ablation, hyperparameter studies, and a case study of our proposed module.

\subsection{Experiment Setup}
\textbf{Datasets.} First, we construct datasets using the CSI100 and CSI300 indices of China's stock market with details in Table~\ref{tab:datasets_details}. Next, we extract a set of 42-dimensional features, which includes 25-dimensional market performance features such as opening price, closing price, change percentage, volatility, and turnover rate, 12-dimensional company valuation features such as P/E ratio, P/B ratio, and P/S ratio, 4-dimensional company categorical features, and 1-dimensional institutional consensus expectations feature. All of these features are normalized prior to analysis.

\begin{table}[!ht]
    \centering
        \small
    \begin{tabular}{l|*{4}{c}}
    \toprule
        ~      & \#\ stocks & \#\ banks &  \#\ industries  & \#\ edges \\ \hline
        CSI100 & 100  & 196 & 97  & 18,950,706   \\ \hline
        CSI300 & 300  & 202 & 191  & 62,500,988  \\ 
    \bottomrule
    \end{tabular}
    % \vspace{-0.1cm}
    \caption{Detailed statistics of the datasets.}
    \label{tab:datasets_details}
\end{table}

\noindent\textbf{Backtest.} We use the timeframe from 01/01/2020 to 02/31/2023 for backtest. The training cycle is set at half a year, meaning that we train the model every six months, resulting in a total of seven models in the experimental set cycle. The training set utilizes data and labels from the preceding six months, while the validation set employs those from last month. The model utilizes fixed parameter values for prediction during the following six months.

\begin{table*}[ht]
    \centering
    \small
    \begin{tabular}{ l | *{4}{c} | *{4}{c}}

        \toprule
       \multirow{2}{*}{Methods} &\multicolumn{4}{c|}{CSI 100}  &\multicolumn{4}{c}{CSI 300}  \\     
          % \specialrule{0em}{2pt}{0pt}
        \cmidrule{2-9} 
        & IC & IR & CR & Prec@30  & IC & IR & CR & Prec@30 \\
        \midrule
         \specialrule{0em}{2pt}{0pt}
        \multirow{2}{*}{MLP}  
&0.0027 &0.0282&0.1166 &0.4751  &0.0039 &0.0314 &0.1721 &0.4958 \\ &(2.25e-03)&(2.26e-02)&(8.10e-03)&(8.17e-04)&(9.42e-04)&(1.53e-02)&(1.08e-02)&(1.01e-03)
        \\
        \specialrule{0em}{2pt}{0pt}
        \cline{1-9}
        
        \specialrule{0em}{2pt}{0pt}
        \multirow{2}{*}{LSTM}  
        &0.0040 &0.0335 &0.1289 &0.4808 &0.0049 &0.0345 &0.1859 &0.4958 \\
        &(1.27e-03)&(1.31e-02)&(1.90e-03)&(2.70e-04)&(6.84e-04)&(1.09e-02)&(1.29e-02)&(1.99e-03)
        \\
        \specialrule{0em}{2pt}{0pt}
        \cline{1-9}

	\specialrule{0em}{2pt}{0pt}
        \multirow{2}{*}{Transformer} 
        &0.0058 &0.0422 &0.1383 &0.4987 & 0.0063 &0.0442 &0.2122 &0.5065 \\
& (2.50e-03) & (1.51e-02) &(7.47e-02)&(3.22e-03) & (1.95e-03) & (1.28e-02) & (1.14e-01)  & (6.03e-03)
          \\
        \specialrule{0em}{2pt}{0pt}
        \cline{1-9}
	
        \specialrule{0em}{2pt}{0pt}
        \multirow{2}{*}{GAT} 
        &0.0031 &0.0274  &0.1534 &0.4812 &0.0066 &0.0454 &0.2653 &0.4991 \\ &(9.08e-04)&(7.63e-03)&(2.45e-02)&(2.31e-03)&(1.50e-03)&(2.46e-02)&(2.42e-02)&(3.00e-04)
        \\
        \specialrule{0em}{2pt}{0pt}
        \cline{1-9}
                
        \specialrule{0em}{2pt}{0pt}
        \multirow{2}{*}{GCN} 
        &0.0038 &0.0305&0.1616 &0.4927&0.0075 &0.0674 &0.2816 &0.5055 \\ &(1.34e-03)&(9.36e-03)&(8.64e-03)&(2.33e-03)&(9.85e-04)&(3.80e-02)&(2.88e-02)&(2.04e-03)
        \\
        \specialrule{0em}{2pt}{0pt}
        \cline{1-9}
        
        \specialrule{0em}{2pt}{0pt}
         \multirow{2}{*}{RGCN} 
         &0.0104 &0.0578  &0.1912 &0.4985&0.0090 &0.0845 &0.5159 &0.5104 \\ &(1.29e-03)&(7.47e-03)&(2.84e-02)&(2.59e-03)&(1.69e-03)&(1.42e-02)&(5.32e-02)&(1.85e-03)
         \\
        \specialrule{0em}{2pt}{0pt}
        \cline{1-9}

	 \specialrule{0em}{2pt}{0pt}
       \multirow{2}{*}{HAN}   
       &0.0108 &0.0525 &0.2267 &0.4997&0.0086 &0.0848 &0.3511&0.5112 \\ &(4.08e-04)&(2.69e-03)&(2.48e-02)&(3.25e-03)&(4.68e-03)&(4.53e-02)&(5.72e-02)&(4.53e-03)
       \\
        \specialrule{0em}{2pt}{0pt}
        \cline{1-9}

        \specialrule{0em}{2pt}{0pt}
       \multirow{2}{*}{HGT}  
       &0.0112 &0.0657& 0.2384&0.5036  &0.0115 &0.0874 &0.4108 &0.4923 \\ &(1.35e-03)&(7.46e-03)&(1.98e-02)&(4.72e-03)&(2.05e-03)&(1.17e-02)&(5.65e-02)&(6.93e-03)
       \\
        \specialrule{0em}{2pt}{0pt}
        \cline{1-9}

        \specialrule{0em}{2pt}{0pt}
       \multirow{2}{*}{EvolveGCN} 
       &0.0065 &0.0538 &0.1815 &0.4961&0.0080 &0.5012 &0.4989&0.4830 \\ &(3.54e-04)&(3.18e-03)&(2.81e-02)&(2.26e-03)&(3.46e-04)&(4.69e-03)&(6.09e-02)&(3.11e-03)
       \\
        \specialrule{0em}{2pt}{0pt}
        \cline{1-9}

        \specialrule{0em}{2pt}{0pt}
       \multirow{2}{*}{HTGNN} 
       &0.0118 &0.0724  &0.2643 &0.5039&0.0192 &0.1773 &0.4653 &0.5126 \\ &(3.76e-03)&(2.45e-02)&(8.23e-02)&(3.54e-03)&(7.59e-04)&(9.94e-03)&(7.03e-02)&(1.12e-03)
       \\
        \specialrule{0em}{2pt}{0pt}
        \cline{1-9}
        \specialrule{0em}{2pt}{0pt}
         \multirow{2}{*}{$\textbf{MDGNN }$}
         & $\textbf{0.0123 }$ & $\textbf{0.0746 }$ & $\textbf{0.2741}$ &$\textbf{ 0.5081 }$  & $\textbf{0.0322 }$ & $\textbf{0.2488 }$ & $\textbf{0.9828 }$  & $\textbf{0.5232}$  \\ 
& $\textbf{(2.75e-03)}$ & $\textbf{(1.59e-02)}$ & $\textbf{(8.11e-02)}$ &  $\textbf{(3.22e-03)}$ & $\textbf{(2.43e-03)}$ & $\textbf{(4.19e-03)}$ & $\textbf{(1.13e-02)}$ & $\textbf{(3.01e-03)}$ 
         \\   
       \bottomrule
    \end{tabular}
     % \vspace{-0.6cm}
     \caption{Results of methods on public datasets. The last row in each dataset indicates the percentage of improvements gained by the proposed method w.r.t the best-performed baseline. Prec@k is a shortened form of Precision@k.} 
    \label{tab:main_results} 
\end{table*}

\noindent\textbf{Baselines.} 
To show the performance of our proposed model, we compare MDGNN with SOTA methods. We select the following models as the baseline for comparison: 
(1) Traditional time series modeling methods (MLP, LSTM~\cite{lstm}, Transformer~\cite{devlin2018bert}). 
%leverage only the intrinsic characteristics of the stock and do not consider the accompanying graph information.
(2) Homogeneous graph methods (GCN~\cite{gcn}, GAT~\cite{gat}).
%solely utilize relationships between stocks and do not incorporate other types of multiplex relationship information.
(3) Heterogeneous graph methods (RGCN~\cite{rgcn}, HAN~\cite{han2023machine}, HGT~\cite{hgt}). 
%exploit heterogeneous relationships but lack the capability to mine dynamic relationships.
(4) Dynamic graph methods (EvolveGCN~\cite{evolvegcn}, HTGNN~\cite{htgnn}). %incorporate the mining of dynamic graph relationships as a fundamental aspect of their approach.

\noindent\textbf{Metrics.}
We utilize Information Coefficient (IC)~\cite{10.1145/3292500.3330833}, Information Ratio (IR), Cumulative Return (CR), and Precision@K~\cite{10.1145/3130348.3130374} as evaluation metrics. 
IC evaluates the overall ranking performance, and IR divides the excess return of a portfolio by its tracking error.
CR is the accumulated portfolio return based on the prediction score.
Precision@K evaluates whether the excess returns of TopK stocks outperform the benchmark index.
%We also calculate the Information Ratio (IR) by dividing the excess return of a portfolio by its tracking error.
%We also calculate IR, that is, the mean of IC divided by the standard deviation of IC. 
%In addition, we constructed a Top N portfolio of daily positions based on model scoring and counted the CR(cumulative return) during the backtest period.

\noindent\textbf{Implementation Details.} 
Our experiment is trained with Nvidia V100 GPU, and all models are built using PyTorch. 
The hidden size was set to 128, the number of GNN layers is 2, and the window size is 10. The training and validation sets are kept consistent across all models.
To ensure that all models receive sufficient training, we train each for 500 epochs and implement an early stopping strategy. 
%with patience of 20. 
%This approach is adopted to ensure thorough training and optimal performance for each model.

\subsection{Experiment Result}
The results of our proposed method, as well as the other baseline models, are presented in Table~\ref{tab:main_results} for CSI100 and CSI300 datasets. Our model outperforms all other methods across all metrics. Based on these experimental findings, we draw the following conclusions:

1) \textbf{Time series modeling and static graphs}:
Traditional time series modeling methods, such as MLP, primarily rely on the intrinsic node features of the stock, while LSTM/Transformer places greater emphasis on temporal features. However, homogeneous graph-based approaches, such as GAT and GCN, only consider node features and stock connections. Despite their inferior performance on the CSI100 dataset, these methods outperform Transformer on the larger CSI300 dataset. 
%This observation underscores the significance of establishing connections between stocks.

2) \textbf{Heterogeneous and dynamic graphs}:
The inclusion of diverse heterogeneous graph information in algorithms, such as RGCN, HAN, and HGT, has led to notable performance enhancements over prior approaches. Additionally, we compared time series heterogeneous graph-based methods such as EvolveGCN and HTGNN, which are designed for temporal and multi-relational graphs, respectively. Our findings suggest that the incorporation of both temporal and multi-relational graph information can yield further improvements in performance.

3) \textbf{Our proposed method}:
Our proposed MDGNN algorithm, leveraging enhanced modules to capture information from the distinctive multi-relational graph structure of stocks, surpasses previous time series heterogeneous graph-based algorithms on both datasets. Moreover, the performance improvement is more pronounced in the CSI300 dataset compared to the CSI100 dataset. This outcome can be attributed to the inclusion of additional institutional and industry nodes, which results in a larger training graph and enables more effective information propagation. These findings provide further evidence of the effectiveness of constructing graphs for stock trend prediction.

\begin{figure*}[ht]
% \vspace{-0.3cm}
\centering
\includegraphics[width=1.6\columnwidth,scale=0.9]{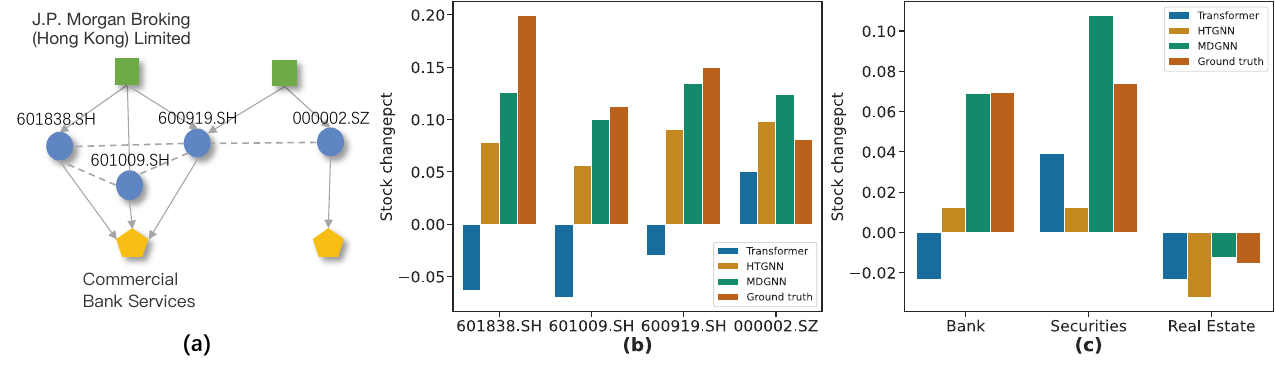}
% \vspace{-0.4cm}
\caption{The results of the case study.}
\label{fig:analysis}
 % \vspace{-0.4cm}
\end{figure*}
 
\subsection{Ablation Study}
\noindent\textbf{Effect of Components.} 
To validate the design choices in our proposed framework, we perform an ablation experiment by removing four components individually: edge weight (w/o edge), meta-path (w/o meta-path), hierarchical aggregation (w/o aggregation), and temporal extraction layer (w/o temporal).
The experiment is performed on the CSI300 dataset, and the results are presented in Table~\ref{tab:ab_result}. We observe that the removal of the meta-path module results in the most significant decrease in performance, thereby confirming the effectiveness of the multi-relational graph in our framework.

\begin{table}[!ht]
    \centering
        \small
    \begin{tabular}{l|*{4}{c}}
    \toprule
        ~ & IC & IR & CR  & Prec@30 \\ \hline
        w/o edge & 0.0268  &0.2155 &0.8950  & 0.5152   \\ \hline
        w/o meta-path & 0.0216  & 0.1723 &0.7502  & 0.5076  \\ \hline
        w/o aggregation  & 0.0303 & 0.2392 & 0.9402 & 0.5227 \\ \hline
        w/o temporal & 0.0286 & 0.2226 & 0.8745 & 0.5215 \\ \hline
        MDGNN & 0.0322 & 0.2488  & 0.9828 & 0.5232 \\ 
    \bottomrule
    \end{tabular}
    \caption{The results of the effect of components.}
    \label{tab:ab_result}
\end{table}

\noindent\textbf{Effect of Relations.} 
To further confirm the effectiveness of each relationship in our multi-relational graph, we present the results in Table~\ref{tab:rel_result}. Here, \textbf{$\mathcal{S}\mathcal{S}$}, \textbf{$\mathcal{S}\mathcal{B}$}, \textbf{$\mathcal{S}\mathcal{I}$}, and \textbf{$\mathcal{I}\mathcal{I}$} refer to the relationships between stocks and stocks, stocks and investment banks, stocks and industries, and industries and industries, respectively. The default connection between different node types is bidirectional, and if the required edges in the meta-path are removed, the corresponding meta-path will also be removed.
%Our findings indicate that the performance is comparable to EvolveGCN when using only the $\mathcal{S}\mathcal{S}$ edge in the homogeneous graph. 
With the $\mathcal{S}\mathcal{B}$ and $\mathcal{S}\mathcal{I}$ edges, the performance improves to some extent, thereby confirming our basic assumption of constructing a multi-relational graph, namely that the stock price changes of stocks held by the same investment bank or belonging to the same industry exhibit a certain degree of consistency. 
%Moreover, after introducing the $\mathcal{I}\mathcal{I}$ edge, the significance of connecting similar industries is validated, as it is more beneficial for information transmission among stocks belonging to similar industries.
Furthermore, the introduction of the connection between investment banks yields a more significant effect than the connection between industries, as the former brings about greater differences in information when multiple investment banks hold a stock, whereas it can only belong to one industry.

\begin{table}[!ht]
    \centering
    \small
    \setlength{\tabcolsep}{1.5mm}
    \begin{tabular}{*{4}{c} | *{4}{c}}
    \toprule
        $\mathcal{S}\mathcal{S}$ & $\mathcal{S}\mathcal{B}$ & $\mathcal{S}\mathcal{I}$ & $\mathcal{I}\mathcal{I}$ & IC & IR & CR & Prec@30 \\ \hline
        \checkmark & - & - & - & 0.0217 & 0.1727 & 0.7372 & 0.5128  \\ 
        \checkmark & \checkmark & - & - & 0.0264 & 0.2092 & 0.8220 & 0.5203  \\ 
        \checkmark & - & \checkmark & - & 0.0210 & 0.1632 & 0.7133 & 0.5101  \\ 
        \checkmark & - & \checkmark & \checkmark & 0.0217 & 0.1802 & 0.7755 & 0.5134  \\ 
        \checkmark & \checkmark & \checkmark & - & 0.0283 & 0.2300 & 0.9074 & 0.5208  \\ 
        \checkmark & \checkmark & \checkmark & \checkmark & 0.0322 & 0.2488 & 0.9828 & 0.5232 \\ 
    \bottomrule
    \end{tabular}
    
    \caption{The results of the effect of relations.}
    \label{tab:rel_result}
\end{table}

\subsection{Case Study}
A research report on investment bank holdings reveals a rising credit pulse trend from December 2021 to March 2022, accompanied by an increase in the proportion of bank holdings by international investment institutions. To analyze this trend, we focus on Chengdu Bank (601838.SH) and its subgraph, which consists of four stock nodes: Chengdu Bank (601838.SH), Nanjing Bank (601009.SH), Jiangsu Bank (600919.SH), and Vanke A (000002.SZ). The first three stocks belong to the commercial banking service industry and are held by the same institution. The last stock belongs to the real estate industry and is held by a different institution, which also holds both Vanke A and Jiangsu Bank.

Figure~\ref{fig:analysis}(b) shows the average stock change rates for the four selected stocks between January 4 and 17, 2022. Traditional temporal models predict negative change rates for most bank-related stocks, except for 000002.SZ. 
However, the MDGNN model enables the upward trend to propagate through multiple relation graphs, influencing the change rates of all bank-related stocks and leading to an increase in the change rate of 600919.SH. 
%The MDGNN model outperforms other models, demonstrating its strong ability to handle multiple relation graphs. 
Further analysis of the average stock change rates of three key industries during this period is shown in Figure~\ref{fig:analysis}(c).
%These industries include banking, securities, and real estate, comprising 22, 7, and 27 stocks, respectively. 
The findings indicate that the utilization of multiple relational graphs has a more pronounced influence on banks and real estate than securities. 
%This observation can be attributed to the consistent trends observed among stocks within the same industry, as well as the larger number of stocks present within these industries, which facilitates the effective propagation of graph information.

\subsection{Hyperparameter Study}
\begin{figure}[ht]
% \vspace{-0.4cm}
\centering
\includegraphics[width=1.0\columnwidth,scale=1.0]{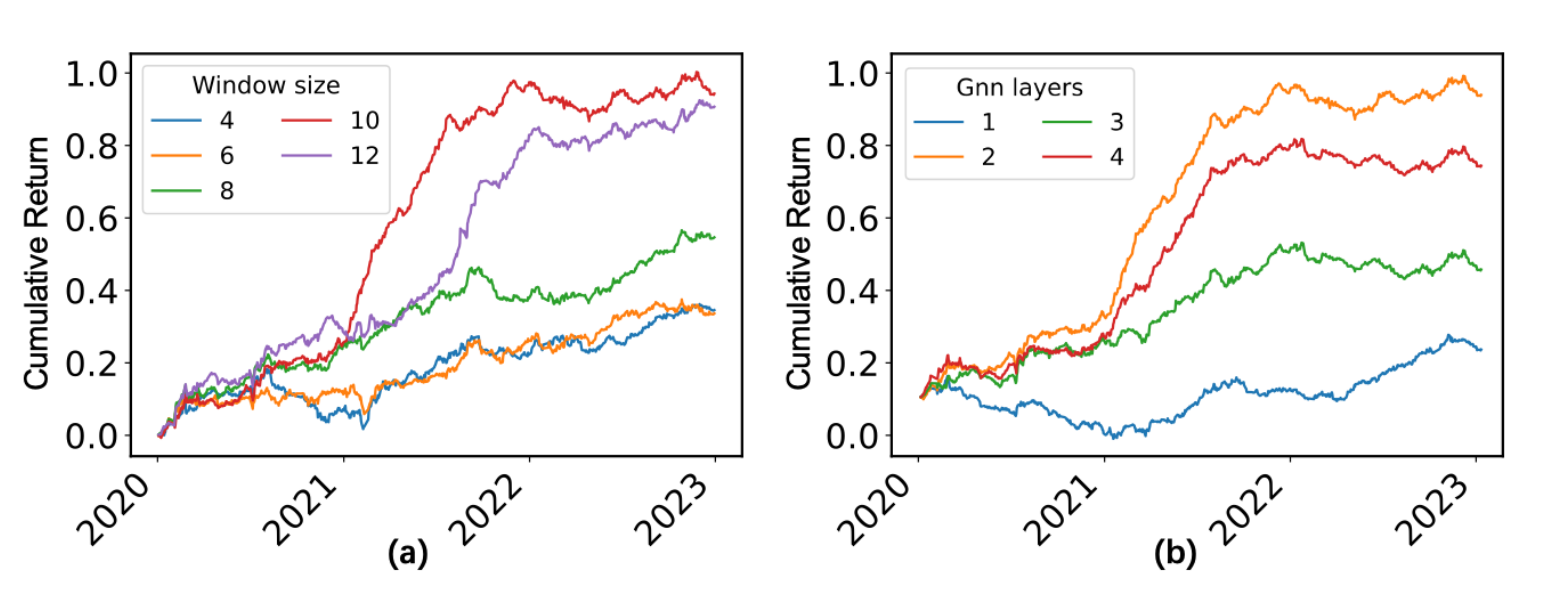}
% \vspace{-0.5cm}
\caption{The results of hyperparameter study.}
\label{fig:comparison}
% \vspace{-0.6cm}
\end{figure}

We also design some experiments to check the sensitivity of hyperparameters. 
In Figure~\ref{fig:comparison}(a), the changes in cumulative return are depicted across different window sizes.
It appears that increasing the window size improves the effect, but only up to a certain limit for information capture. 
Similarly, as the number of GNN layers increases in Figure~\ref{fig:comparison}(b), the effect also improves gradually, but an excessively high complexity can lead to a decline in performance.

%%%%%%%%%%%%%%%%%%%%%%
\section{Conclusion}
In this work, we formally define the multifacetedness and temporal patterns of stocks through empirical analysis for the first time and propose a novel hierarchical multi-relational dynamic graph framework for modeling stock investment prediction.
Our approach involves constructing a multi-relational graph for each trading day and generating a set of discrete graph snapshots within the specified look-back window size. In terms of the intra-day graph snapshot, we design a hierarchical multi-relational graph embedding layer to first aggregate the neighbor nodes within a specific meta-path and then adaptively integrate the stock representation from distinct meta-paths. 
Furthermore, we incorporate the transformer structure to aggregate the temporal evolving patterns of stocks. 
We demonstrate the effectiveness and robustness of our proposed framework through extensive experiments. 
In the future, we would like to study MDGNN with contrastive learning methods for stock investment prediction and improve performance even further.

%%%%%%%%%%%%%%%%%%%%%%
%\newpage

\bibliography{aaai24}

\end{document}